\documentclass{wscpaperproc}
\usepackage{latexsym}
\usepackage{graphicx}
\usepackage{mathptmx}
\usepackage{float}

\usepackage{amsmath}
\usepackage{amsfonts}
\usepackage{amssymb}
\usepackage{amsbsy}
\usepackage{amsthm}
\usepackage[pdftex,colorlinks=true,urlcolor=blue,citecolor=black,anchorcolor=black,linkcolor=black]{hyperref}

\newtheoremstyle{wsc}% hnamei
{3pt}% hSpace abovei
{3pt}% hSpace belowi
{}% hBody fonti
{}% hIndent amounti1
{\bf}% hTheorem head fontbf
{}% hPunctuation after theorem headi
{.5em}% hSpace after theorem headi2
{}% hTheorem head spec (can be left empty, meaning `normal')i

\theoremstyle{wsc}

\begin{document}

%***************************************************************************
% AUTHOR: AUTHOR NAMES GO HERE
% FORMAT AUTHORS NAMES Like: Author1, Author2 and Author3 (last names)
%
%		You need to change the author listing below!
%               Please list ALL authors using last name only, separate by a comma except
%               for the last author, separate with "and"
%

% setting up general page style
\pagestyle{fancyplain}

% setting up page style of first page
\thispagestyle{plain}
\firstPageHead{}

% setting up running header (authors) of subsequent pages
\chead{\fancyplain{}{\itshape Ram\'irez de Los Rios, V\'asquez Correa and Escudero Mar\'in}}

% setting up seperation parameters
%\headsep=72pt
\rhead{}
\cfoot{}
\renewcommand{\headrulewidth}{0pt} % (renewcommand needed in fancyhdr to remove top decorative line)
%\headrulewidth=0pt  % ("setlength" needed in fancyheading to remove top decorative line)

%%%%%%%%%%%%%%%%%%%%%%%%%%%%%%%%%%%%%%%%%%%%%%%%%%%%%%%%%%%%%%%%%%%%%%%%%%%%%%
%                                                                            %
%     THESE COMMANDS ARE REQUIRED TO WORK WITH WSC.BST TO MAKE BIBLIO     %
%                                                                            %
%%%%%%%%%%%%%%%%%%%%%%%%%%%%%%%%%%%%%%%%%%%%%%%%%%%%%%%%%%%%%%%%%%%%%%%%%%%%%%
\makeatletter
\let\@internalcite\cite
\def\cite{\def\@citeseppen{-1000}%
    \def\@cite##1##2{(##1\if@tempswa , ##2\fi)}%
    \def\citeauthoryear##1##2##3{##1 ##3}\@internalcite}
\def\citeNP{\def\@citeseppen{-1000}%
    \def\@cite##1##2{##1\if@tempswa , ##2\fi}%
    \def\citeauthoryear##1##2##3{##1 ##3}\@internalcite}
\def\citeN{\def\@citeseppen{-1000}%
%  Pierre L'Ecuyer's fix for multiple cite bug
%  Added by Paul J Sanchez on 4 October 2001
%   \def\@cite##1##2{##1\if@tempswa , ##2)\else{)}\fi}%
%   \def\citeauthoryear##1##2##3{##1 (##3}\@citedata}
    \def\@cite##1##2{##1\if@tempswa, ##2)\else{}\fi}%
    \def\citeauthoryear##1##2##3{##1 (##3)}\@citedata}
\def\citeA{\def\@citeseppen{-1000}%
    \def\@cite##1##2{(##1\if@tempswa , ##2\fi)}%
    \def\citeauthoryear##1##2##3{##1}\@internalcite}
\def\citeANP{\def\@citeseppen{-1000}%
    \def\@cite##1##2{##1\if@tempswa , ##2\fi}%
    \def\citeauthoryear##1##2##3{##1}\@internalcite}
\def\shortcite{\def\@citeseppen{-1000}%
    \def\@cite##1##2{(##1\if@tempswa , ##2\fi)}%
    \def\citeauthoryear##1##2##3{##2 ##3}\@internalcite}
\def\shortciteNP{\def\@citeseppen{-1000}%
    \def\@cite##1##2{##1\if@tempswa , ##2\fi}%
    \def\citeauthoryear##1##2##3{##2 ##3}\@internalcite}
\def\shortciteN{\def\@citeseppen{-1000}%
%  Pierre L'Ecuyer's fix for multiple cite bug
%  Added by Paul J Sanchez on 2 September 2002
%  should have caught this last year...
%   \def\@cite##1##2{##1\if@tempswa , ##2)\else{)}\fi}%
%   \def\citeauthoryear##1##2##3{##2 (##3}\@citedata}
% Shane G. Henderson fix for extra right bracket at end of optional material June 8, 2005
%    \def\@cite##1##2{##1\if@tempswa, ##2)\else{}\fi}%
    \def\@cite##1##2{##1\if@tempswa, ##2\else{}\fi}%
    \def\citeauthoryear##1##2##3{##2 (##3)}\@citedata}
\def\shortciteA{\def\@citeseppen{-1000}%
    \def\@cite##1##2{(##1\if@tempswa , ##2\fi)}%
    \def\citeauthoryear##1##2##3{##2}\@internalcite}
\def\shortciteANP{\def\@citeseppen{-1000}%
    \def\@cite##1##2{##1\if@tempswa , ##2\fi}%
    \def\citeauthoryear##1##2##3{##2}\@internalcite}
\def\citeyear{\def\@citeseppen{-1000}%
    \def\@cite##1##2{(##1\if@tempswa , ##2\fi)}%
    \def\citeauthoryear##1##2##3{##3}\@citedata}
\def\citeyearNP{\def\@citeseppen{-1000}%
    \def\@cite##1##2{##1\if@tempswa , ##2\fi}%
    \def\citeauthoryear##1##2##3{##3}\@citedata}
%
% \@citedata and \@citedatax:
%
% Place commas in-between citations in the same \citeyear, \citeyearNP,
% \citeN, or \shortciteN command.
% Use something like \citeN{ref1,ref2,ref3} and \citeN{ref4} for a list.
%
\def\@citedata{%
    \@ifnextchar [{\@tempswatrue\@citedatax}%
                  {\@tempswafalse\@citedatax[]}%
}

\def\@citedatax[#1]#2{%
\if@filesw\immediate\write\@auxout{\string\citation{#2}}\fi%
  \def\@citea{}\@cite{\@for\@citeb:=#2\do%
    {\@citea\def\@citea{, }\@ifundefined% by Young
       {b@\@citeb}{{\bf ?}%
       \@warning{Citation `\@citeb' on page \thepage \space undefined}}%
{\csname b@\@citeb\endcsname}}}{#1}}%

% don't box citations, separate with ; and a space
% also, make the penalty between citations negative: a good place to break.
%
\def\@citex[#1]#2{%
\if@filesw\immediate\write\@auxout{\string\citation{#2}}\fi%
  \def\@citea{}\@cite{\@for\@citeb:=#2\do%
    {\@citea\def\@citea{, }\@ifundefined% by Young
       {b@\@citeb}{{\bf ?}%
       \@warning{Citation `\@citeb' on page \thepage \space undefined}}%
{\csname b@\@citeb\endcsname}}}{#1}}%

% (from apalike.sty)
% No labels in the bibliography.
%
\def\@biblabel#1{}
\makeatother

%\newlength{\bibhang}
%\setlength{\bibhang}{2em}

% Indent second and subsequent lines of bibliographic entries. Taken
% from openbib.sty: \newblock is set to {}.
% \renewcommand{\refname}{REFERENCES}

\newdimen\bibindent
\bibindent=0.0em
% SEC: was \def\thebibliography#1{\section*{\refname\@mkboth
% SEC: was   {\uppercase{\refname}}{\uppercase{\refname}}}\list
\def\thebibliography#1{\section*{\refname}\list
   {}{\settowidth\labelwidth{[#1]}
   \leftmargin\parindent
   \itemindent -\parindent
   \listparindent \itemindent
   \itemsep 0pt
   \parsep 0pt}
   \def\newblock{}
   \sloppy
   \sfcode`\.=1000\relax}

           % Set up BiBTeX macros

% needed to make the tex document look more like the word counterpart :-(
\setlength{\baselineskip}{12.7pt}

% AUTHOR: Enter the title, all letters in upper case
\title{ABMS OF SOCIAL NETWORK BASED ON AFFINITY}

% AUTHOR: Enter the authors of the article, see end of the example document for further examples
\author{Juan Camilo Ram\'irez de los Rios\\ [12pt]
Mathematical engineering student\\
EAFIT University\\
jurami28@eafit.edu.co\\
% Multiple authors are entered as follows.
% You may also need to adjust the titlevbox size in the preamble - search for titlevboxsize
\and
Mar\'ia Camila V\'asquez Correa\\[12pt]
Mathemathical engineering student \\
EAFIT Univeristy\\
mvasqu49@eafit.edu.co\\
\and
Paula Alejanda Escudero Mar\'in\\ [12pt]
 Mathematical science department professor \\
EAFIT University\\
pscuder@eafit.edu.co\\
}

\maketitle

\section*{ABSTRACT}

An agent based model is proposed for analyzing the dynamics that rise from interactions within social networks, analyzing the individual behavior of each profile. Said model considers a simplified construction of a social network, while satisfying properties attributed to this type of systems. For that matter, previously established studies on the matter are taken into account, while including each profiles preferences, and how they evolve with time, for the system's dynamic behavior. Results are analyzed based on concepts like the emergence of clusters, the polarization of the network and the homogeneity of preferences between connected profiles; and how they may depend on the characteristics of the profiles and of the network.
\section{INTRODUCTION}
\label{sec:intro}

The main factors that determine the dynamic behavior of social networks are its profiles, the interactions between them and their characteristics \cite{Girvan}. In fact, social networks are systems based mainly on proactive, adaptive and autonomous individuals. Furthermore, their individual behavior and decision making process can be modeled, while aggregated dynamics are not so straightforward to comprehend. Because of this, agent based modeling and simulation is a compelling option to study social networks, and the conditions under which interesting properties may emerge.

For this, a simplified version of a social network is presented, and is the base for the presented model and its analysis. Said simplification is a modified version of previously analyzed social network models, which satisfies characteristics commonly attributed to those type of systems, like low overall connectivity, high emergence of clusters and the presence of many profiles with few connections and few with many connections. Furthermore, links between profiles are created, maintained or modified based on the similarities in preferences between them. Then, the main consideration for the behavior regarding connections has to do with the characteristics of each profile. Finally, various simulations are generated for different conditions of the network and its individuals, and outputs like the emergence of clusters, degree of polarization of the network, homogeneity of preferences between profiles and the network's degree of connectivity are analyzed. Results, conclusions and recommendations are then mentioned.

\section{ODD Protocol}
\label{sec:ODD}
\subsection{Overview}

\subsubsection{Purpose}

To observe the dynamics rising from the interactions within a social network, regarding the degree of polarization, clustering and connectivity within it. At first, the network will be composed as presented by \cite{tote}, while adding new decision rules. 
For considering this extra criteria, it will be assumed that all interactions in the network happen regarding a certain and unspecified ideology. For instance, everyone on the network could be focused on politics. These new dynamics may change the structure of personal networks and, in extension, the whole social network. \\

\subsubsection{Entities, State Variables and Scales}

Individuals/Social network profiles and links between them are the entities considered in this model. Each profile has various state variables: age, survival rate, affinity, sensibility and influentiability.
Furthermore, links are directed and have a state variable, the strength/level of the connection, which is not necessarily reciprocal; it may be different between the link from profile A to profile B, and the one from B to A.\\ \ \\
Also, the length of one time step in the simulation is specified as 10 days, and simulations will consist of 1,000 time steps (10,000 days). In addition, the total amount of members of the social network is given by a variable, \textit{maxProfiles}, as provided by a slider.

\subsubsection{Process overview and scheduling}

Each time step, the model executes the following processes in the described order:

\begin{enumerate}
\item Create new connections for each profile.
\item Update each profile's network.
\item Update each profile's affinity. 
\item Decide randomly which profiles are going to be erased from the social network. 
\item Replace erased profiles
\end{enumerate}

\subsection{Design concepts}

\subsubsection{Basic Principles}
  The basic principles considered are the construction of a personal network as described by \cite{tote} in which an individual's connections can be separated into five levels:
\begin{enumerate}
\item Strongest: closest relatives and a few close friends.
\item Strong: emotionally important close friends and relatives with whom
relationships are maintained.
\item Medium: emotionally important close friends and relatives with whom
relations are passively maintained.
\item Weak: people who are important for “economic and social purposes
and the logistics of everyday life”.
\item Weakest: acquaintances, whose names may not be known.
\end{enumerate}
 
The construction of each personal network will follow the aforementioned principles, however, as the analysis is for Internet-based social networks, physical distance between agents is not taken into account. The things considered in order to change each personal network are the interests individuals have in common with each other \cite{wtf}; specifically, how the behaviour of an individual is affected by the ideologies of its connected profiles. 
\\
\\
In order to do this, the variable \textit{affinity} is introduced. Affinity reflects a particular individual's opinion regarding the network's discussed topic. Said variable is measured from 0 to 1.This is used to generate two ideological extremes (0 and 1), and places in between.
\\ \ \\
To aid in the modification of personal networks based on affinities, which is more thouroughly explained in section 2.5, the following concepts are introduced:
\begin{enumerate}
\item Sensibility: The probability that someone rejects a person that has a different point of view from theirs.
\item Influentiability: the proportion from the change in affinity generated by an individual's close connections, that is actually changed. For instance, if the proposed change is $change$ and the individual's influentiability is $inf$, then the modification to $aff$ will be of $inf * change$.
\end{enumerate}

It is important to note that the nature of the ideology considered, regardless of the topic it is about, is not taking into account in the model. Anyhow, its nature can be reflected trough the sensibility and influentiability variables. For instance, for topics regarding politics and sports, it is arguable that people tend to be more stubborn than when dealing with music or the environment. That is, it is easier for someone to reject a friend if their politic position is completely different, but harder if we are talking about music taste. 

\subsubsection{Emergence}
The emergence of communities and outliers is evaluated. Communities being groups of connected profiles which have few connections to other groups, and outliers being profiles which have very big or very small personal networks.

\subsubsection{Adaptation}
The adaptive behavior of profiles is represented by variations on their personal networks: removal, addition or modification of connections with other profiles, as well as changes in their affinity. This is modeled by some decision rules that emulate the real evolution of personal networks: profiles relating with profiles of similar interests. 
\subsubsection{Objectives}
The main objective of each individual is to keep their personal network composed by profiles they like, where liking is understood as having a similar affinity level; while also trying to be related to people of a similar age. In order to do this, the individual will keep his closest circles composed of his favorite profiles, and his less liked ones will be amongst the weakest connections. Furthermore, each social circle, and each personal network as a whole will have to be kept within a certain size, as specified by parameters.
\subsubsection{Learning}
The model does not consider any kind of learning.
\subsubsection{Prediction}
Profiles in the model do not consider information in order to make predictions.
\subsubsection{Sensing}
Each profile can sense the affinity and age of their connected profiles, and also the strength of their connection. Additionally, based on the level of the connection, the perception of the other profile's variables is distorted: the strongest the connection, the more accurate the perception is. No other state variables are sensed. 
\subsubsection{Interaction}
Profiles interact with other profiles with which they have a connection, and in turn to the connection between them. This interaction is necessary for the variation of the affinity variable of each profile. Also, links interact with the profiles they are connecting in order to establish their strength level, or to be eliminated altogether. Finally, two unconnected profiles may establish a connection between them.  

\subsubsection{Stochasticity}
\label{sec:stoc}
Initialization variables are generated as follows: 

\begin{itemize}
\item Although this does not seem to be the case in real social networks, age is set from an uniform distribution between 10 and 80, for simplicity purposes.
\item Affinity is generated from an uniform distribution between 0 and 1, assuming opinions on the topic are varied and equiprobable.
\item Sensibility is also a random uniform number between 0 and 1, assuming no prior knowledge about each individual's psychological reactions to disagreement on the topic being study.
\item Influentiability is also from a uniform 0 1.
\end{itemize}
Furthermore, individuals have a probability to "die" at any moment, which means being removed from the network. This is called the survival rate. That is, if in a certain time, a random number is lower than an individual's survival rate (from an exponential distribution on age), then the person dies. This is a modified process, described originally by \cite{tote2}. It is important to note that this idea represents human survival rates fairly accurately.
\\ \ \\
Moreover, the perception of a connected profile's affinity is distorted by a normal distribution with mean 0, and a certain standard deviation. The latter being set by the user by a slider (with values from 0 to 0.2). The variable with the given deviation will distort the perception of the affinity of the \textit{strongest} connections and, for every connection of lower importance, the deviation of the normal distribution will be multiplied by 1.1. This aids in showing that profiles know other profiles better the more closely related they are.

\subsubsection{\ Collectives}
No collectives are explicitly modeled. Regardless of this, studies over real social networks, like those presented by \cite{Girvan}, indicate that communities of accounts with similar interests emerge. These communities have also been showed to strongly influence the interests of their members. They appear and are studied throughout the model, but they are not modelled as collectives. 
\subsubsection{\ Observation}
Observations are based on various measures of the model: 

\begin{itemize}
\item Network density: the amount of connections active at the moment for the whole network, standardized by the amount of total possible connections.
\item Average personal network's size: the average amount of links going out from a profile.
\item Affinity: the opinion regarding the topic which is being discussed in the social network.
\item Amount of outliers: the amount of profiles whose personal networks' size is very small and also those with very big networks.
\item Clustering coefficient: a representation of how clustered the network is.
%\item Short path lengths: the length of the shortest path between two given profiles.  
\end{itemize}
These variables are used to evaluate how connected the social network is, and to identify some of the characteristics of said connections.

\subsection{Details}

\subsubsection{Initialization}
Here, we describe the value (or default value) of the parameters of the model:
\begin{itemize}
\item Number of people: The default value is 100
\item Maximum number of individual connections: The default value is 50
\item Maximum number of links of each type. Calculated with the following percentages of the maximum number of individual connections
\begin{itemize}
\item Strongest: 5\%
\item Strong: 10\%
\item Medium: 20\%
\item Weak: 30\%
\item Weakest: 40\%
\end{itemize}
\item Distortion: default value 0.05.
\item Max change: the maximum amount the affinity is allowed to change in each iteration. Default value: 0.15 
\item Affinity radius: The value at which a profile keeps a connection with another one or the connection changes its strength (or disappears). That is, given the affinity of a profile, the profiles that are in the range [affinity - affinity radius, affinity + affinity radius] keep or create new connections. default value of 0.2.
\item Number of people to die at each time step. Default value of 5.
\end{itemize}
As described above, the setup of the model will give every person a random age, a random affinity, sensibility and influentiability and no connections to other people.
\subsection{Input Data}
There's no input data for model.
\subsection{Submodels}
\label{Sub}
We present in detail each process of the model:

\begin{enumerate}

\item Search for new connections: if the personal network size is not at the limit, the person looks up for potential connections from profiles with which no directed link has been established, and selects the $n$ closest profiles in terms of the variable $age$, where $n$ is the minimum between the 30\% of $maxNetwork$ and the amount of individuals left to fill his weakest level of friends. Then, those profiles within $affinity$ values of $affinity-radius$ are selected. 
\\

\item Evaluate each profile's network: profiles can sense, albeit  partially, the affinity of other's profiles to the ideology, and the level of the connection they have with said profile. The perceived value, which is the real value with some random Gaussian distortion, as was previously explained, along with the relationship strength, will be kept for later. This process happens to every profile, without any specific order. 
\\
\\
Then, they modify their network based on the collected perceived affinity. For each other connected profile, if the overall difference between their affinity and the individual's own is above parameter \textit{aff-radius}, a random number is generated and compared to the $sensibility$ value of the original profile. Afterwards, the experiment can indicate two things: if the random number exceeds $sensibility$, do nothing; otherwise, reduce the strength of the connection with the other profile. If the link was of type "weakest", then the connection would be eradicated. Furthermore, if the perceived affinity is within a certain range (a fraction of $affinity-radius$, calculated based in link strength), then the strength of the link will increase. More precisely, given an own affinity $aff$ and an affinity-radius of $ar$, the strength of a link will be defined if the perceived affinity of the other profile is within a certain range:

\begin{itemize}
\item $\text{Weakest links for } \left [ aff - ar,\ aff + ar \right ]$ 
\item $\text{Weak links for } \left[ aff\ - \frac{4}{5} ar,\ aff +\frac{4{}}{5} ar \right]$ 
\item $\text{Medium links for } \left[ aff\ - \frac{3}{5} ar,\ aff +\frac{3}{5} ar \right]$ \item $\text{Strong links for } \left[ aff\ - \frac{2}{5} ar,\ aff +\frac{2}{5} ar \right]$ \item $\text{Strongest links for } \left[ aff\ - \frac{1}{5} ar,\ aff +\frac{1}{5} ar \right]$ 
\end{itemize}

\item Affinity variations: affinity values for linked profiles are used to calculate a variation in the value for that same variable, but of the profile itself. As such, a weighted average between the perceived affinities(which are the real affinities with a Gaussian noise) of connected individuals, with a link strength equal or superior to medium, and the own affinity value, is calculated, giving more priority to those coming from more closely connected accounts. Then, this average is multiplied by the influentiability parameter from the original profile. Finally, as the value can change at most \textit{max change}, either the affinity value is changed for the newly calculated value, or it is modified by \textit{max change}.
 \\

\item Decide if each agent is going to be removed from the model: as described in \ref{sec:stoc}, with the survival rates given (that are generated by an exponential distribution and expected to increase with age) and a random number, the death of an agent may happen. The parameter $peopleDead$ controls the maximum amount of profiles which will drop from the model each time step. Furthermore, it is important to note that if a profile reaches age $80$, it will immediately be removed from the simulation.
\\
\item Replace removed agents: this process is also described in \ref{sec:stoc}, a dead person is replaced by a new one (as suggested by \cite{tote2}), with age 10, random influentiability, affinity and sensibility, and no connections.
\end{enumerate}

\section{Background}
\label{sec:back}
Two very enlightening papers on the matter of studying social networks through agent based modelling and simulation are: \cite{tote} and \cite{tote2}. These papers present some traditional models used for describing social networks in literature, and also show some of the problems these have to replicate with precision the real behavior of social networks. For that matter, a different approach is presented, considering the following characteristics of social networks:

\begin{itemize}
\item Be of limited size.
\item Display high clustering.
\item Change over time.
\item A low overall personal network density.
\item Presence of communities.
\item Presence of a few profiles with many connections, and many with few connections.

\end{itemize}

Regardless of this, the aforementioned papers consider a concept called "social distance" for personal network modification. For instance, two profiles will create connections if they are socially close to each other, and the magnitude of the proximity will also affect the strength of the link. Anyway, they do not consider social network dynamism, as it is generated but not modified trough time. In this paper though, a slight modification of the model presented in \cite{tote} is initialized, while the network is modified dynamically due to drop outs from the model, changes in affinity and the stochasticity involved in estimating the affinity of other profiles.

\section{Implementation}
\begin{figure}[h]
	\centering
    \includegraphics[scale = 0.5]{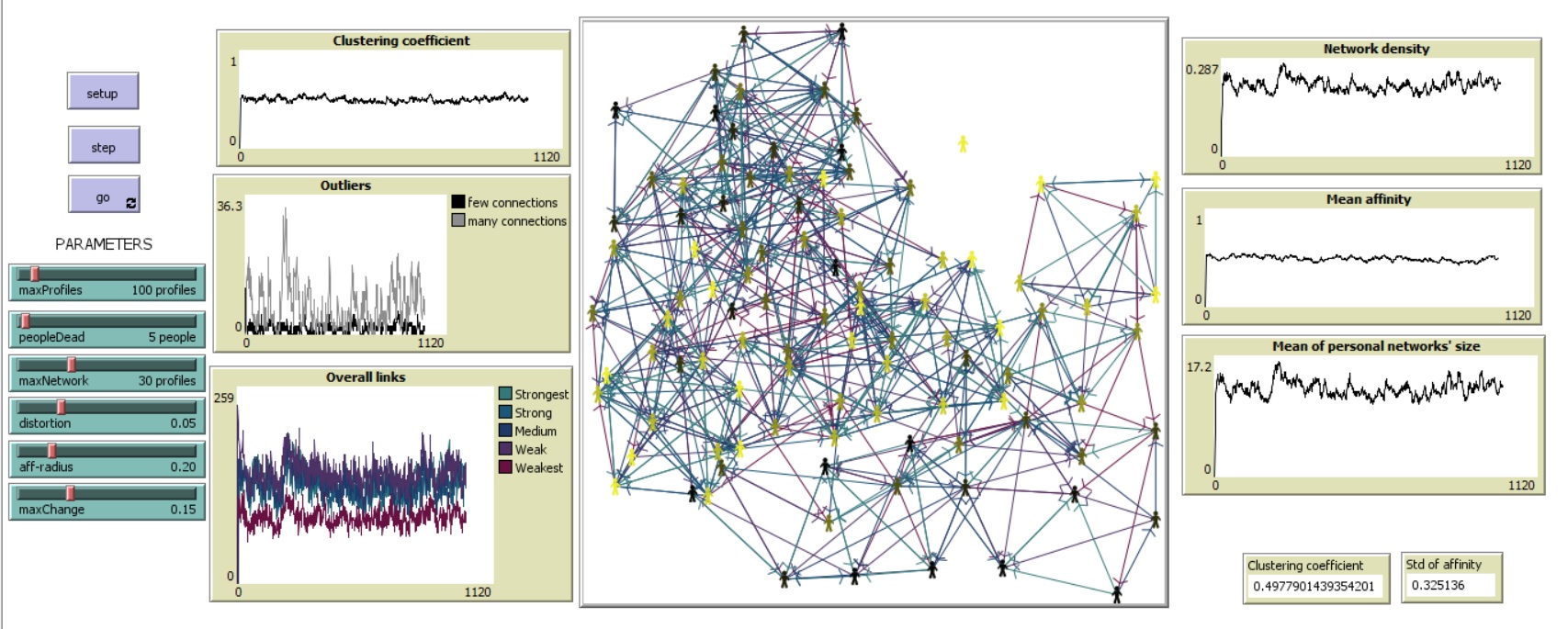}
    \caption{Results for the default value of parameters}
    \label{fig:def100}
\end{figure}

The implementation of the model was made on NetLogo, following the process overview described above. Now, given the default values of the parameters, we show the graphics of the model at $t = 1000$ (the end of the simulation, Figure \ref{fig:def100}): \\

There exist some design decisions taken for the sake of the visualization, such as:

\begin{itemize}
\item There are two monitors: Clustering coefficient and Standard deviation of affinity. The first one was introduced to observe the exact value of this output at the end of the simulation and the second one is only a monitor cause we are interested in observing the change of the original standard deviation at the end of the simulation
\item Graphics show:
\begin{itemize}
\item The clustering coefficient over the time
\item The outliers: In gray, the ones that have many connections (almost the limit of the connections that can be made) and in black, the ones that have very few connections
\item The network density: We expect this value to remain low, so we follow it over the time
\item Mean affinity: This and the standard deviation give some information about the polarization of the population
\item Mean of personal network's size: This ant the information about outliers allow us to determine if the connections made are uniform or have a lot of deviation
\item Overall links: Each color reflects the color o the actual link in the space shown, and we expect to track the strength of the connections made in the network
\end{itemize}
\item There are sliders for all the parameters that are not generated randomly
\item The graphical representation, in the center, shows the profiles as people colored by the level of their affinity. That is, the yellow ones (the lighter color) are the ones with 0 affinity and the black ones (the darkest color) are the ones with 1 affinity. The rest of the colors can be interpreted in consequence. \\
Also, links are colored by their strength. The color of each one can be visualized in the \textit{Overall links} graphic. 
\end{itemize}
In the info tab of the model are specifications about the use and the functioning of the whole model, as well as some recommendations for extending it. The \textit{SocialNetwork.nlogo} is attached to this document.
\section{Verification and Validation}
\subsection{Verification}
Extreme conditions aid in the verification of the model. For instance, Figure \ref{fig:exp1}, which represents the case where parameter $aff-radius$ is $0$ (which means profiles will only relate themselves to other profiles with the \textit{same} affinity), no connections are generated throughout the simulation. Furthermore, for an opposite value of $1$, where everyone wants to friends with everyone else, $network\ density$ rises to $0.671$, which represents very high levels of connectivity, as is shown in Figure \ref{fig:exp2}. It does not reach a value of $1$, though, due to the need of profiles to maintain connections with other profiles with similar ages, which thereby causes rejection towards some members of the social network. \\
\\
Also, as can be seen in Figure \ref{fig:exp5}, where the affinity variable is allowed to change at most 0 in each iteration of the simulation, the mean value of said variable approaches $0.5$ (its expected value coming from an uniform distribution). Apart from this, the simulation reaches stationarity fast, and manages to maintain it for long periods of time, due to no drastic changes in the network's environment as people only die of age. Also, there is an interesting visualization of various clusters, which contain similar colored agents, which represent profiles of similar affinities. For a deeper insight into these topics, consider viewing Section \ref{sec:popper}.

\subsection{Validation}
No real social network is considered by this paper, but rather a simplification of one. Because of this, no data is available for validating the model in relation to a real system, which does not exist altogether. Regardless, the model is designed to replicate some of the results of \cite{tote} and it should satisfy principles attributed to social networks. These are presented in Section \ref{sec:back}.
\\
\\
For instance, by construction and due to parameter $maxProfiles$, the considered network is limited in size. Furthermore, Figure \ref{fig:def100} indicates relatively high clustering, through a clustering coefficient of $0.46$; which gets even higher for other values of the parameters, like those presented in Figure \ref{fig:exp5}. Also, Figure \ref{fig:exp5} aids for the graphical display of the formation of communities. Additionally, outliers with very many connections are low, and those with very few as well. Finally, the Network density, with coherent values for the parameters displays low values, around $0.2$.

\section{Results}
\subsection{Extreme values of parameters}
In this section, we discuss some results obtained by varying some of the parameters to an extreme point. The point of the parameter space is specified in each experiment. As usual, the simulation is for $10000$ days and the results are displayed bellow. \\
For this, it is useful to analyze the results of the simulation with the default value of the parameters, obtained in Figure \ref{fig:def100}
Now, let's consider the following points from the parameter space and analyze its outputs:
\begin{enumerate}
\item A trivial scenario comes with the $aff-radius = 0$. As shown in Figure \ref{fig:exp1}, no links are created, due to the fact that it is highly improbable to have someone with exactly the same affinity.
\begin{figure}[h]
	\centering
    \includegraphics[scale = 0.55]{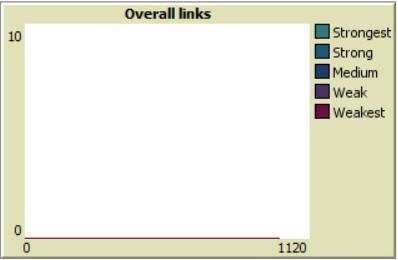}
    \caption{Results for $aff-radius = 0$}
    \label{fig:exp1}
\end{figure}
\item Now, we can consider the $aff-radius = 1$, that is, they will accept everyone that is in the range of age, keeping the size of the network. Results are shown in Figure \ref{fig:exp2}.
One can notice the high network density, which means that there exist a lot of links relative to all the possible ones. In the graphic for the overall links, one can note that the highest ones are the medium and strong links, while the weakest and strong are a few and, as expected for the percentage permitted of the maxNetwork parameter, the strongest links are the fewer ones. This shows the fact that people create connections with a considerable strength.\\
\\
Let us now look at the $affinity$ variable. While its mean holds around $0.5$, which means that, overall, the community of profiles has not gone towards either extreme($0$ or $1$) of said variable. But also, the standard deviation of $affinity$ is $0.38$ and that indicates the network is somehow polarized. Note that each $affinity$ value comes from an uniform distribution between $0$ and $1$, which has a theoretical standard deviation of approximately $0.29$. Because of this, there is information for indicating that profiles have gone to either extreme, which represent polarization. Furthermore, the clustering coefficient is over $0.5$, that means communities are raising. \\
\begin{figure}[h]
	\centering
    \includegraphics[scale = 0.45]{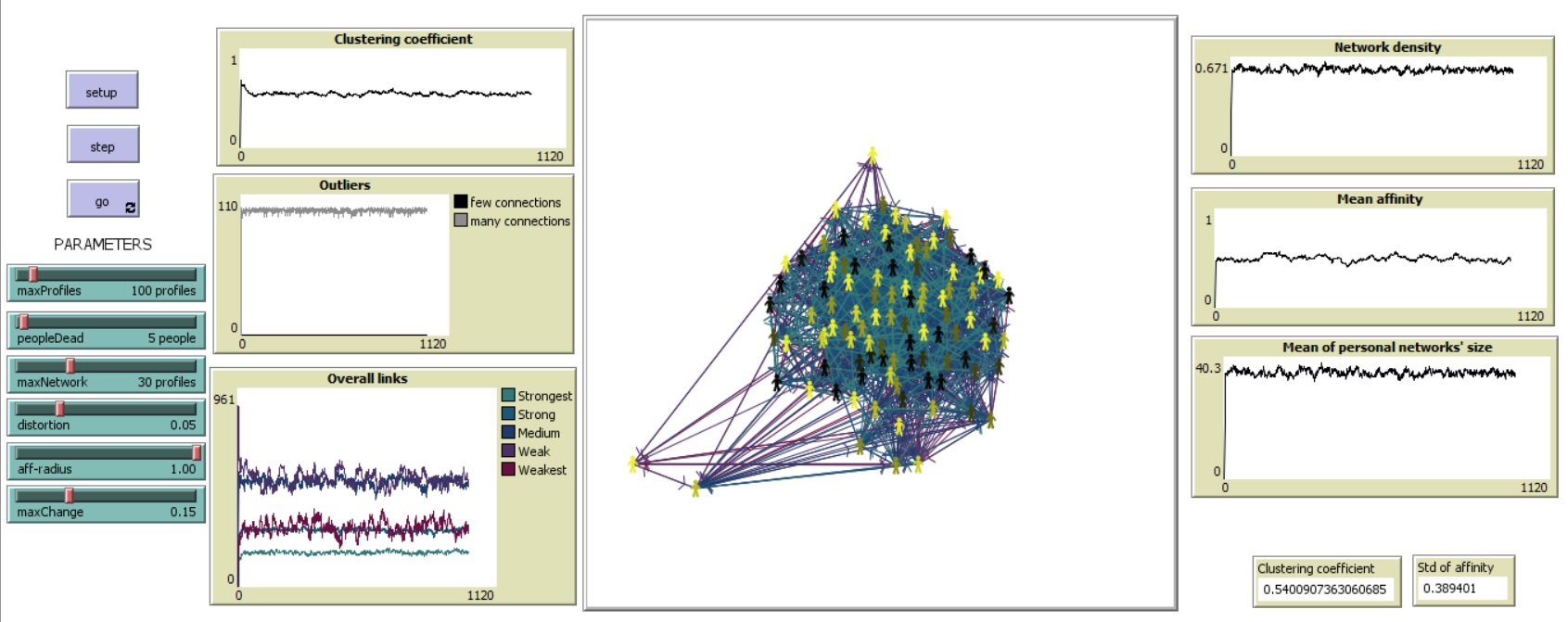}
    \caption{Results for $aff-radius = 1$}
    \label{fig:exp2}
\end{figure}
\item Now, let's consider the case when no changes to the personal affinity are allowed to happen. In the first one, shown in Figure \ref{fig:exp4}, the other parameters remain stationary. Anyway, the mean affinity keeps changing due to the fact that 5 random people are dying at each time step. The clustering coefficient is high, $0.557$, and the standard deviation is close to the distribution of the generator of the affinity. \\

On the other hand, let's consider a network in which people only die when they reach the top age. This simulation is shown in Figure \ref{fig:exp5}. This is a peculiar scenario, in which we can observe the characteristics of social networks. However, people do not get influenced, so they are not prone to generate polarization. And, observing the cluster coefficient of $0.6$, it can be determined that clusters are emerging. Looking at the graphics, one can conclude the same: clusters exist and are made by people with similar affinities, as represented in the different colors of the people in the simulation\\

\begin{figure}[h]
	\centering
    \includegraphics[scale = 0.45]{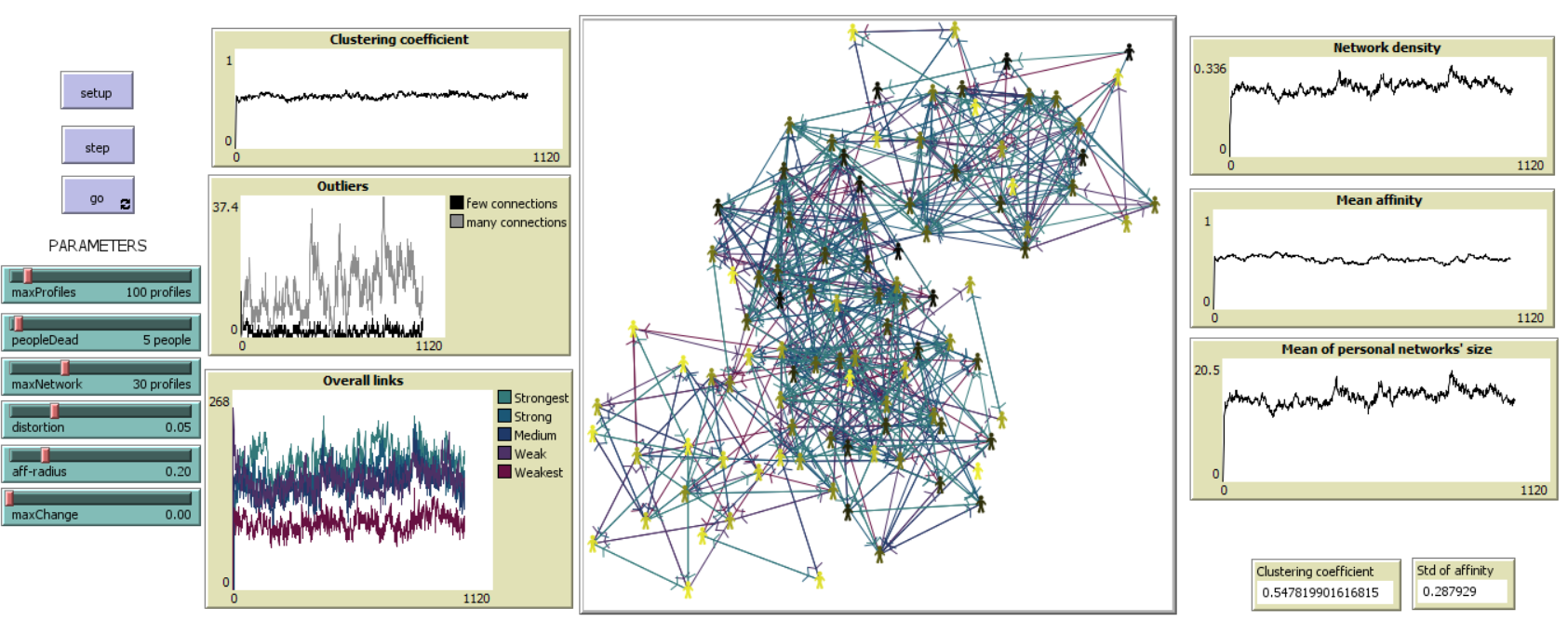}
    \caption{Results for $maxChange = 0$}
    \label{fig:exp4}
\end{figure}
\begin{figure}[h]
	\centering
    \includegraphics[scale = 0.45]{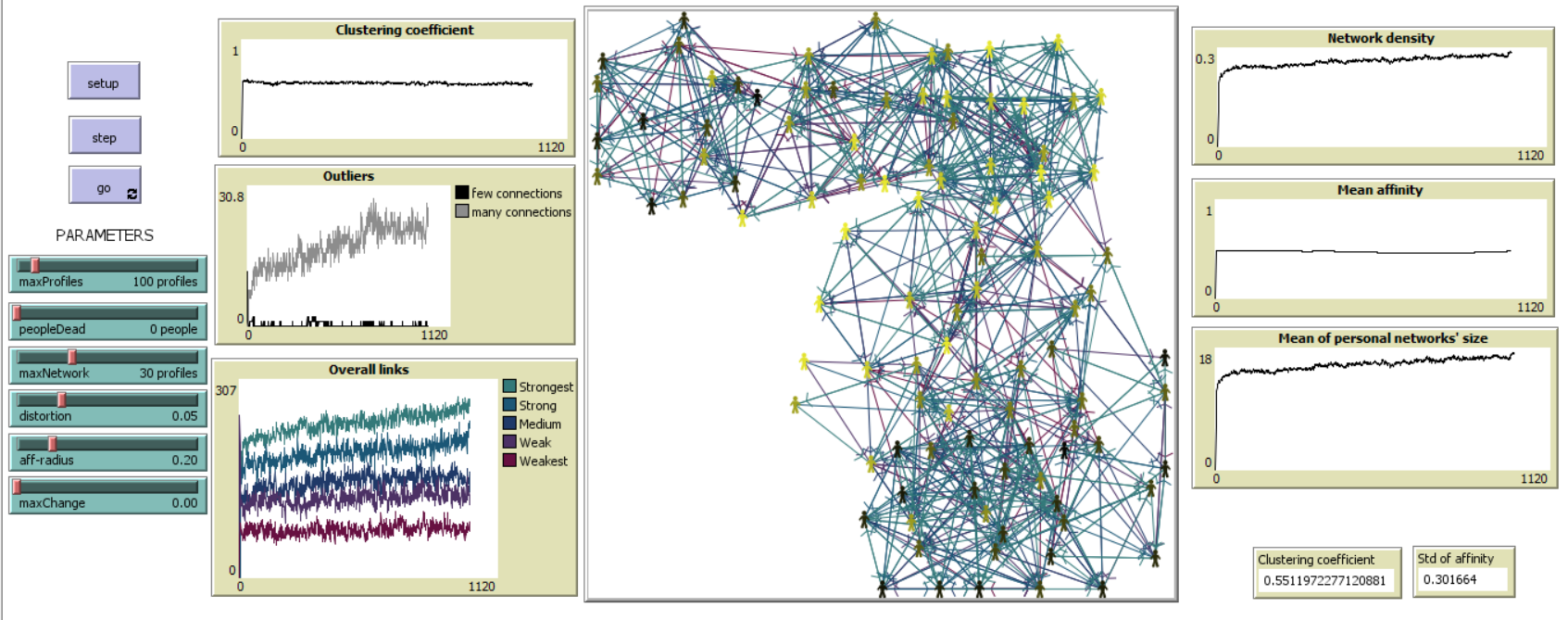}
    \caption{Results for $maxChange = 0$ and $peopleDead = 0$}
    \label{fig:exp5}
\end{figure}
\item Adjusting the $maxNetwork$ parameter to the maximum possible ($maxProfiles - 1$), we can see, in Figure \ref{fig:exp6}
\begin{figure}[h]
	\centering
    \includegraphics[scale = 0.45]{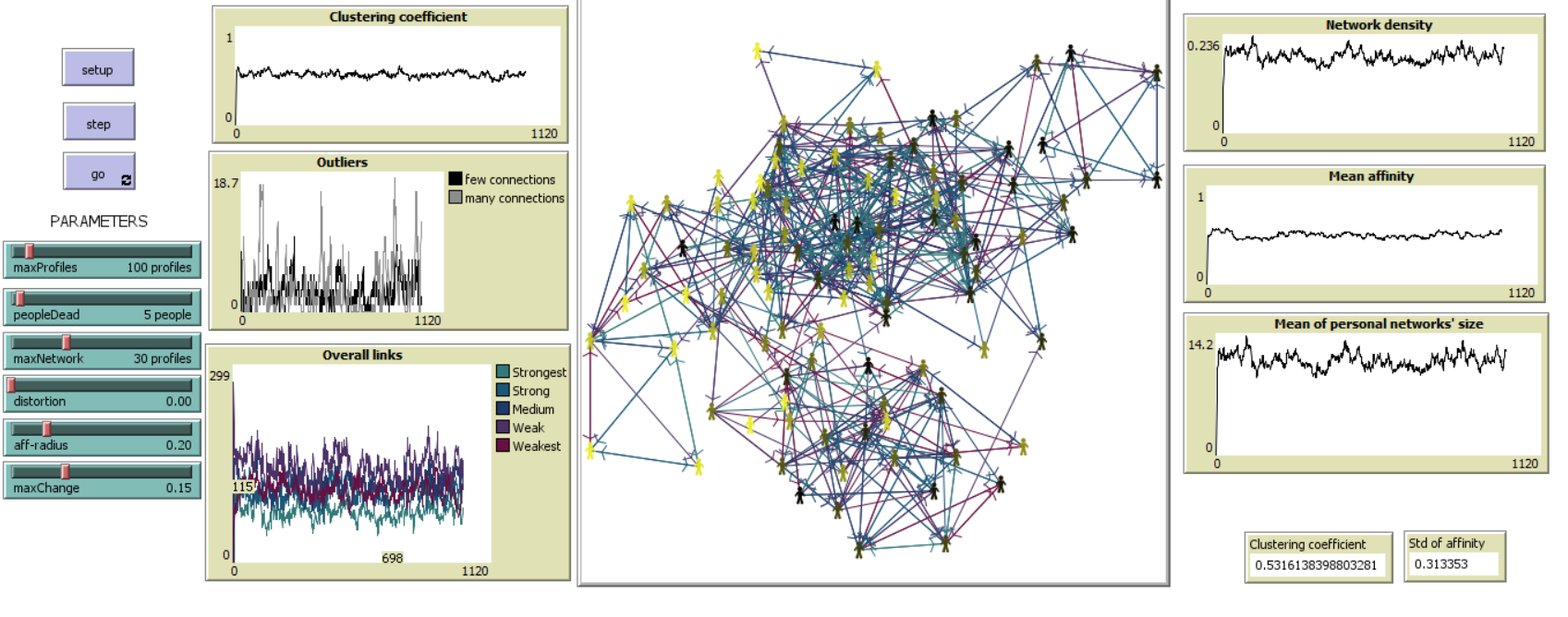}
    \caption{Results for $maxNetwork = 99$}
    \label{fig:exp6}
\end{figure}
\item Another interesting parameter is the $distortion$, Let's analyze the visual outputs for the $distortion = 0$ and $distortion = 1$ that we obtain.\\
For the first one, looking at Figures \ref{fig:exp7} and \ref{fig:exp8}, we observe various things: In the first experiment, with default parameters, even when there exists the possibility of a change in the affinity, the standard deviation remains close to the expected form an uniform distribution, and the mean fluctuates, probably because of the people dying in each tick. And in the second case, the mean affinity stays almost the same trough time and we observe, again, the forming of communities or clusters with similar affinities, fact that indicates polarization in the network. 

\begin{figure}[h]
	\centering
    \includegraphics[scale = 0.45]{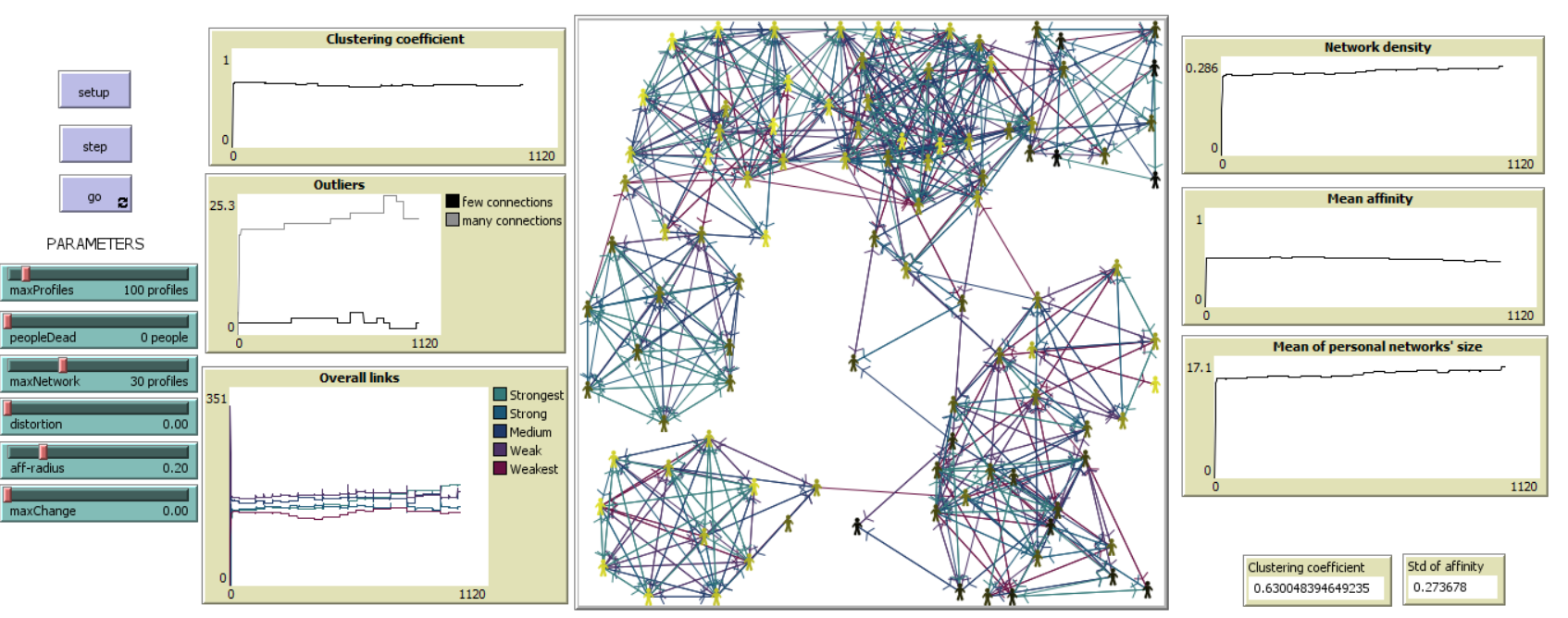}
    \caption{Results for $distortion = 0$}
    \label{fig:exp7}
\end{figure}
\begin{figure}[h]
	\centering
    \includegraphics[scale = 0.45]{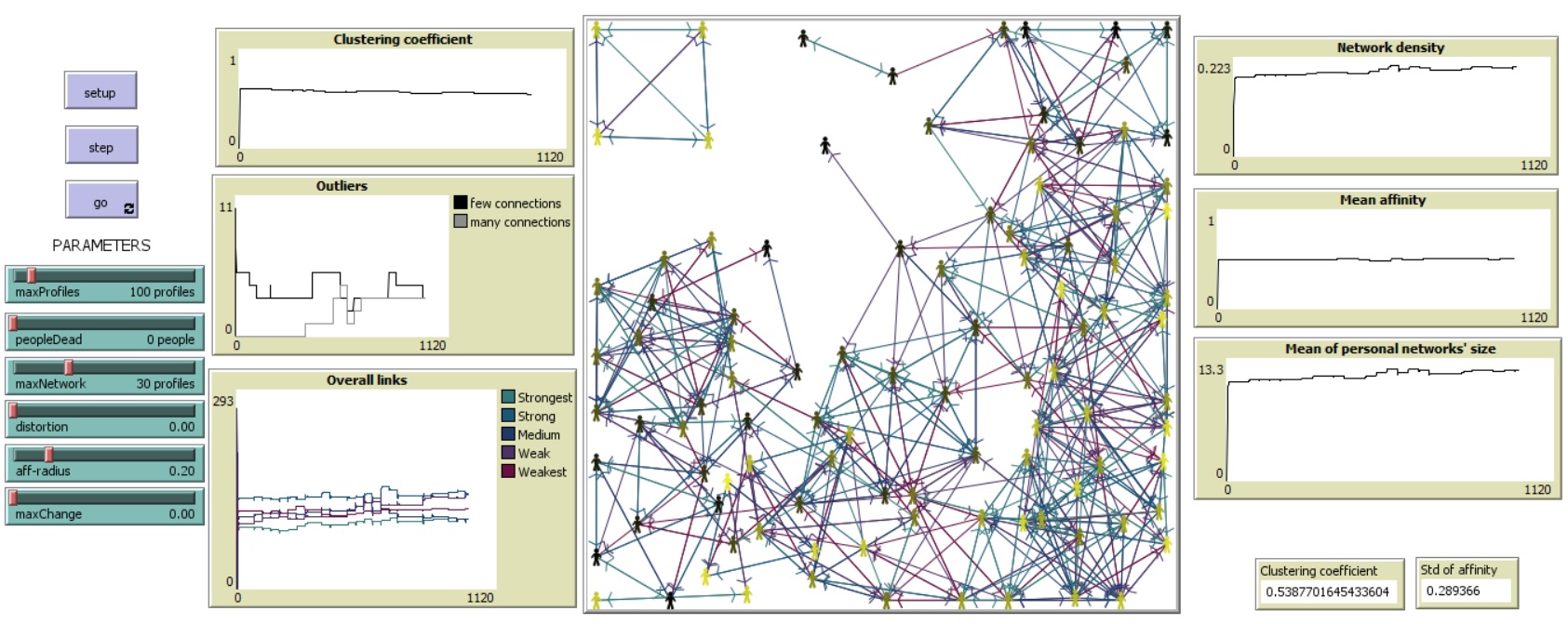}
    \caption{Results for $distortion = 0$, $peopleDead = 0$, $maxChange = 0$}
    \label{fig:exp8}
\end{figure}
\end{enumerate}
\subsection{Experiments over parameters}
One of the most important outputs for our model is the clustering coefficient. That is the reason we perform experiments to see how the cluster coefficient changes if the parameters change. Consider, the point of the parameter space that is the default described and the changes made in each of the following parameters. We show the experiment designed an a graphic with the clustering coefficient at every one of the configurations made:
\begin{enumerate}
\item \textit{maxChange}: The range for changing this parameter was from 0 to 1 with a step of 0.2. Results for the experiment are shown in Figure \ref{fig:expmc}. While the maxChange for the affinity is low (it starts at zero), the clustering coefficient reports high values, that meaning that people tend to form more communities or to have more stable ones that grow over time if their affinity isn't changing trough time. And, when the max change approaches 1, this clustering coefficient descends (though it stays relatively high), indicating that, the fact of being greatly influenced by the people in your social network makes it more difficult for the people to form communities. 
\begin{figure}[h]
	\centering
    \includegraphics{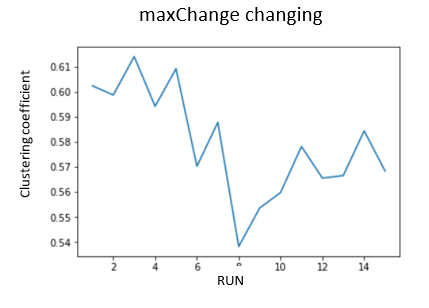}
    \caption{Results for the maxChange experiment}
    \label{fig:expmc}
\end{figure}
\item \textit{aff-radius}: 
Varying the parameter from 0.1 to 0.5 with a step of 0.05, we obtained the results shown in Figure \ref{fig:expar}. In the first 5 runs, a really low clustering coefficient can be seen. This can be considered as an unrealistic scenario, because it is unlikely for people to only be related to very similar people. However, it can seen that the formation of clusters or communities with this setup of parameters is not very high. Now, for the last runs that correspond to the higher values for $aff-radius$, the clustering coefficient rises, meaning that allowing a lot of different profiles in their network makes it easier to form clusters.
\begin{figure}[h]
	\centering
    \includegraphics{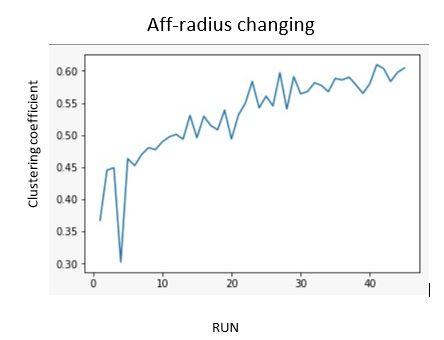}
    \caption{Results for the aff-radius experiment}
    \label{fig:expar}
\end{figure}
\item \textit{maxNetwork}: The range for changing this parameter was 10 to 99 (the maximum permitted) with a step of 10. The results obtained are shown in Figure \ref{fig:expmn}. As expected for the definition of clustering coefficient, the fact that people can create more connections generate more communities. Also, low values for this parameter result in a unrealistic scenario, where the cluster coefficient does not approach the one for a social network. 
\begin{figure}[h]
	\centering
    \includegraphics{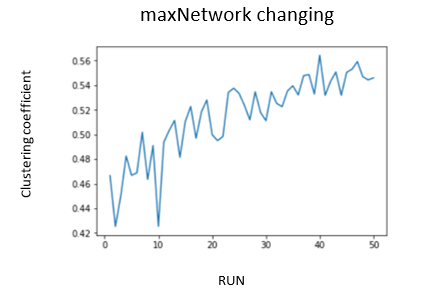}
    \caption{Results for the maxNetwork experiment}
    \label{fig:expmn}
\end{figure}
\end{enumerate}
\subsection{Currencies of the model outputs}

There are various outputs from the model which aid in the analysis of its behavior:

\begin{itemize}
\item Mean affinity plot: displays the mean affinity of all profiles at each tick of the simulation. A value very separate from 0.5 indicates the network is biased, and a centered one may represent either polarization or homogeneity. It is expected that under realistic values for the parameters, this value will stay around 0.5. 

\item

Standard deviation affinity monitor: shows how close together all the profiles are. For instance, a low value for the monitor indicates homogeneity, and a high one may represent a polarized network. It is expected that if the affinity radius is small, and maxChange high, then the value for this output will be high, there will be high clustering, and each cluster will have very similar individuals in terms of affinity.

\item Network density plot: displays the network density on every time step. Lower values indicate an overall lower connectivity. Because of that, it is expected that higher values of $aff-radius$ will be reflected in higher values for this output, as profiles tend to get connected to each other more easily.

\item Clustering coefficient plot and monitor: represents the clustering coefficient, as calculated in \cite{tot}. Here, it is expected that if people change less their affinity in each iteration, then it is more likely that small groups of profiles are generated, which are not heavily modified over time. Also, a simulation with lesser distortion might present higher clustering than one with lots of uncertainty, due to personal network's being constantly modified, and allowing, due to perception errors, connections with profiles with very different affinities which may not me connected between them.

\item Outliers plot: plots the amount of profiles with social network size 10\% or smaller (black) and 70\% or greater (gray), which represents very connected and very unconnected people. 
\item Overall links plot: presents the total amount of links of each kind in the whole networks, and aids in comparing how tightly connected profiles are. 

\end{itemize}
It is also expected that, if the sensibility variable for each profile were generated from a higher mean random variable, then polarization would be higher, as rejection might happen more commonly.

\subsection{ Analyze from the bottom up}
\label{sec:popper}

Procedures will be analyzed regarding their functionality in relation to how they should behave:

\begin{itemize}
\item 

If the model is initialized with $10$ agents created with the same age and affinity, and 9 maximum connections, then running $create-connection$ with parameter $9$, every profile becomes connected to every other profile, as is expected because their ages and affinities are close enough. For random ages and affinities, only some links, at most 9, are be created.

\item
Creating two turtles with very similar affinity values, while maintaining $distortion$ within a reasonable magnitude, and then moving tick after tick of the simulation, their connection will become of the strongest type. Similarly, by starting with two turtles different in affinity, they will eventually destroy the link between them. 

\item Setting various profiles, all connected to each other profile, while they all have an affinity of 1, except for one which has a value of 0, and letting the simulation run, makes the affinity of the different turtle to increase very drastically, and the other turtles face a slight decrease of said variable.

\item

Also, when initializing the model with $10$ individuals, and a $peopleDead$ value of $0$, then calling the $kill$ procedure, even if repeatedly, for each turtle, nothing happens. If $peopleDead$ is modified for $5$, each time the procedure is called, at most 5 people die.

\item
Then, simply calling $replace-agents$ as an observer generates new agents, to complete a total number of 10. Also, these new profiles have no connections and age 10. (ask turtles [show age] \& ask turtles [show count my-out-links])

\end{itemize}

\subsection{ Explore unrealistic scenarios}
An unrealistic scenario is the case when all turtles created with an affinity of 0. Here, the clustering coefficient of the network is very close to 1, and there are no changes in affinity for any profile. Here, when new individuals enter the network, they are slowly attracted towards the others and end up having low affinities; or, they form a cluster of high affinity profiles.

\section{Sensitivity analysis}

The standard deviation of the affinity variable and the cluster coefficient of the network will be evaluated for sensibility analysis, using 30 runs of the model. The value of these outputs for the default value of the parameters is: \\
Clustering coefficient: 0.492862640052283	
\\
Standard deviation of affinity : 0.315100570262851
\\
\\
For each of the following parameters, the effect of either a 5\% or 10\% variation will be evaluated trough the aforementioned outputs, obtained the mean with a total of $20$ runs for each setup of parameters. Then, the sensitivity coefficient will be evaluated.

\begin{figure}[h]
	\centering
    \includegraphics[scale = 0.6]{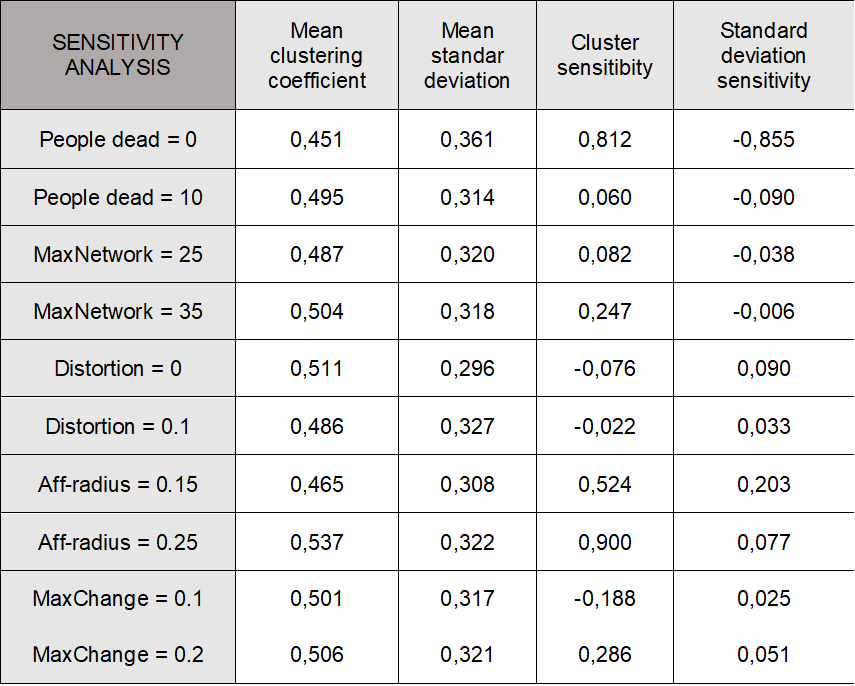}
    \caption{Results for the sensitivity analysis performed}
    \label{fig:sens}
\end{figure}

Outputs are very sensitive to the changes in some parameters. For instance, a lower people dead parameter causes pretty substantial changes in the values of both the cluster coefficient of the network and the standard deviation of the affinities. This can be due to the difficulties for the network to stabilize and maintain links when people are constantly dying and being replaced by new people with different characteristics in each time step. Also, an increase in said variable is not as significant because it represents a maximum quantity of dead agents, not a fixed numbers. \\

Furthermore, aff-radius also generates changes in the monitored variables, especially cluster coefficient, probably due to the possibility of more interconnection betweenfiles to a profile. different connected profiles to a profile. \\

In comparison with the other parameters, maxNetwork and maxChange are very sensible when talking about clustering coefficient. As discussed above, if people are allowed to create more connections, it is expected for them to find more similar people and to stick to this configuration of the network. Similarly, if the change in the affinity is not very high, one can see that people will keep the networks they formed over time, making them more similar their own affinity.

\section{Conclusions and Recommendations}

\subsection{Conclusions}

Ultimately, social networks are very complex systems because of the intervention of many profiles which present different behaviors. Regardless of this, agent based modeling and simulation, while considering personal preferences as a means to explain interaction within social networks, is a capable tool. Said techniques indicate that these systems tend to form clusters and can get very polarized when its profiles are more stubborn regarding their preferences, but it can also reach high levels of connectivity if said profiles allow change. By all counts, the structure and characteristics of a social network and its profiles has great influence in its dynamics. 
\\ 

\subsection{Recommendations}

This model presents a simplified version of a social network and its reach is slightly limited when trying to analyze real networks with a lot of very different profiles. However, it develops various concepts regarding social behavior and could be extended in the following ways:
\begin{itemize}
\item Make the affinity radius a variable for each profile, not a parameter. This makes sense since not every people is willing to accept in their network the same difference in affinity. This is somehow reflected in the sensibility variable and thus could be used to calculate each profile's affinity radius.
\item The model does not consider mutual connections with other profiles, nor the longevity of a link when it is about to change it. Furthermore, at the moment of creating new connections, it would be interesting for profiles to consider other profiles with which thay share many mutual connections.
\item Also, to specify an ideology, as mentioned before, one could modify data from particular affinities and sensibility, that reflect how important the position of others about a topic affects the real population. It can also be done by obtaining the affinity and sensibility from a different distribution, not the uniform one. 

\end{itemize}

% Please don't exchange the bibliographystyle style
\bibliographystyle{wsc}
% AUTHOR: Include your bib file here
\bibliography{demobib}
\end{document}